\documentclass{iau} 
\usepackage{graphicx}

\title[DAGAL] 
{DAGAL: Detailed Anatomy of Galaxies}

\author[Johan H. Knapen]   
{Johan H. Knapen$^{1,2}$}

\affiliation{$^1$Instituto de Astrof\'\i sica de Canarias, E-38205 La Laguna, Tenerife, Spain\\[\affilskip]
$^2$Departamento de Astrof\'\i sica, Universidad de La Laguna, E-38205 La Laguna, Tenerife, Spain\\[\affilskip]}

\pubyear{2015}
\volume{321}  
\jname{ Outskirts of Galaxies}
\editors{A. Gil de Paz, J. Lee, J. H. Knapen, eds.}
\begin{document}

\maketitle

\begin{abstract}

The current IAU Symposium is closely connected to the EU-funded network DAGAL (Detailed Anatomy of Galaxies), with the final annual network meeting of DAGAL being at the core of this international symposium. In this short paper, we give an overview of  DAGAL, its training activities, and some of the scientific advances that have been made under its umbrella. 

\end{abstract}

\firstsection 
\section{Introduction}

One of the EU funding lines of interest for basic research projects in astronomy is that of the Initial Training Networks (ITN, now called Innovative Training Networks). We applied for a network called DAGAL, for Detailed Anatomy of Galaxies, in 2011, and were one of the only 8\% of proposals which were funded. DAGAL had its origin in the {\it Spitzer} Survey for Stellar Structure of Galaxies (S$^4$G; Sheth et al. 2010), with the European co-Is of the S$^4$G project forming the core team of DAGAL. The academic nodes of DAGAL are the Instituto de Astrof\'\i sica de Canarias (IAC, PI and coordinator node), Universidad Complutense de Madrid, the Universities of Groningen, Oulu, and Marseille, and the Max Planck Institut f\"ur Astronomie in Heidelberg, while the private sector partners are Springer (www.springer.com), Fractal (www.fractal-es.com), and Specim (www.specim.fi). 

DAGAL started on April 1, 2012, and the IAU Symposium marks its end (which is formally on 31 March 2016). The total team involved in DAGAL is around 30 people, including 10 fellows (2 postdocs and 8 PhD students) and a part-time project manager at the IAC. We have organised meetings and workshops, run a training programme, done outreach, and produced well over a hundred papers describing scientific advances. We summarise the main achievements of the network below. 

\section{Training and Outreach}

The main aim of the ITN scheme is to provide high-level training to a new generation of European scientists, which they can then apply in their future careers, whether in academia, industrial research, or elsewhere. The training offered within DAGAL was thus a combination of research training in the context of a PhD or postdoctoral research programme, and training in complementary skills. The latter includes meeting organisation, presenting, scientific and outreach writing, project management, software use, and language training. This training was delivered to the fellows by their supervisors or team members, by their own universities, or in specific training courses organised by the DAGAL network. For instance, a customised project management course was delivered by our private sector partner Fractal during one of the annual meetings.

Network-wide research training activities were also offered, and included the use of large telescopes (all fellows participated in an observing run on the 2.5m Isaac Newton Telescope on La Palma), training schools on numerical modelling and radio and IFU astronomy, and lectures by invited experts during the annual network meetings.

From the outset, DAGAL has stimulated public outreach by its team members and fellows. As a result, the project has delivered press releases, radio programmes, and a number of ``pretty pictures'' of galaxies on its webpage (http://www.dagalnetwork.eu/image-gallery). The fellows maintained the DAGAL twitter and facebook accounts, with 500 followers, and another fellow won a prize for an outreach article he wrote in his home country. All fellows gave talks to the general public or to school children, in their home countries, place of work, or elsewhere.  

The DAGAL network organised three annual meetings, three training schools, a Lorentz Center workshop in October 2014, and now, expanded from its fourth and final annual meeting, an international scientific conference which has achieved IAU Symposium status and which has been organised by DAGAL's Madrid node in the beautiful city of Toledo.

\section{Public data products}

One of the main products of the DAGAL project is a set of scientific deliverables, which are publicly available (http://www.dagalnetwork.eu/public-scientific-deliverables and http://www.astro.rug.nl/$\sim$dagal/). They include an advanced search and sample selection tool and links to all reduced S$^4$G {\it Spitzer} images, {\it GALEX} UV and optical images of most of them, and H{\sc i}, CO, or H$\alpha$ kinematic and imaging data for subsets. Also available are an atlas of structures in S$^4$G galaxies (such as bars, lenses, and rings) and a catalogue of gas flows in barred potentials: gas and orbit responses to typical potentials, with varying axial ratio, pattern speed, and bar mass.

\section{Scientific Results}

To date, over 130 refereed papers (see http://www.dagalnetwork.eu/publications) are associated with the DAGAL project, of which over 50\% have attracted more than three times the median number of citations in our field. Rather than reviewing all of these new results, I will briefly highlight some of the work published by the fellows in the DAGAL network. For instance, Leaman et al. (2015) identified a merging dwarf galaxy companion to the galaxy NGC~7241, and deduced from dynamics, stellar populations and star formation analyses that this is a merger event with a mass ratio similar to the Milky Way - Small Magellanic Cloud pair, with an initial passage some 2\,Gyr ago. 

Querejeta et al. (2015) used independent component analysis to separate the 3.6 and 4.5\,$\mu$m S$^4$G images into emission from old stars and dust emission, which allows us to derive accurate stellar mass maps (publicly available on http://www.astro.rug.nl/$\sim$dagal/). These maps were used to derive that interactions increase the star formation rate in galaxies, but only by a factor of around two (Knapen, Cisternas \& Querejeta 2015). R\"ock et al. (2015) used infrared spectra to extend empirical stellar population models to the mid-IR, vital for the interpretation of star formation histories in galaxies (see also R\"ock et al. 2016; for the models see http://miles.iac.es)).  Bouquin et al. (2015) presented a {\it GALEX}/S$^4$G $UV-IR$ colour-colour diagram and deduced that galaxies quickly transition from the so-called {\it GALEX} blue to the red sequence, on timescales of order $10^8$\,yr.

Galactic bars in the S$^4$G sample were characterised in detail by D\'\i az-Garc\'\i a et al. (2016), uncovering possible evidence for the growth of bars within a Hubble time from the density amplitudes, lengths, and strengths of bars in galaxies of different morphological types. Herrera-Endoqui et al. (2015) created a catalogue of morphological features in the complete sample of the S$^4$G, reporting among many other parameters the normalised sizes of components like inner rings and lenses. One conclusion is that barlenses indeed form part of the bar. Using the same images from the S$^4$G in combination with high-quality H{\sc i} data,  A. Ponomareva et al. (in prep.) study the intrinsic scatter in the Tully-Fisher relation, uncovering a high sensitivity of the scatter to the internal kinematics of the gas. 

Tsatsi et al. (2015) discovered how prograde mergers of two galaxies can produce kinematically decoupled cores (KDCs) in the resulting elliptical galaxies, adding another powerful mechanism to form the substantial number of observed KDCs in early-type galaxies. The imprints of boxy/peanut structures on the kinematics of simulated disk galaxies was studied by Iannuzzi \& Athanassoula (2015), while Fragkoudi et al. (2015) explored the effects of modelling a boxy/peanut bulge on the estimates of bar galaxy parameters. Fragkoudi et al. found that failing to allow for a boxy/peanut bulge can induce errors of up to 40\% in the forces in the bar region, and significantly impacts the study of a modelled galaxy's orbital structure, kinematics and morphology. 

\section{Conclusions}

DAGAL was an EU-funded training network which ran from April 2012 to March 2016. Its final annual meeting formed the core of IAU Symposium~321, and provided a good opportunity to review and celebrate its successes. The network has employed eight PhD students and two postdocs, all of whom have made successful steps in their career development. We encourage our colleagues to use the DAGAL data products which are publicly available, and which include images as well as catalogues of advanced results. 


\section{Acknowledgments}

The author wishes to thank all those who have helped make DAGAL a success: the fellows, the staff members, the DAGAL manager Alejandra Mart\'\i n, and many others. We acknowledge support to the DAGAL network from the People Programme (Marie Curie Actions) of the European Union's Seventh Framework Programme FP7/2007-2013/ under REA grant agreement PITN-GA-2011-289313, and from the Spanish Ministry of Economy and Competitiveness (MINECO) under grant number  AYA2013-41243-P. JHK thanks the Astrophysics Research Institute of Liverpool John Moores University for their hospitality, and the Spanish Ministry of Education, Culture and Sports for financial support of his visit there, through grant number PR2015-00512.

\end{document}